\documentclass{article}
\usepackage{spconf,amsmath,graphicx}
\usepackage{booktabs}
\usepackage{amssymb}
\usepackage{bm}
\usepackage{multirow}
\usepackage{arydshln}
\usepackage{cite}
\usepackage{tabularx}
\usepackage{url}
\usepackage{setspace}
\usepackage[breaklinks=true]{hyperref}

\DeclareMathOperator*{\argmax}{argmax}

\makeatletter
\def\adl@drawiv#1#2#3{%
        \hskip.5\tabcolsep
        \xleaders#3{#2.5\@tempdimb #1{1}#2.5\@tempdimb}%
                #2\z@ plus1fil minus1fil\relax
        \hskip.5\tabcolsep}
\newcommand{\cdashlinelr}[1]{%
  \noalign{\vskip\aboverulesep
           \global\let\@dashdrawstore\adl@draw
           \global\let\adl@draw\adl@drawiv}
  \cdashline{#1}
  \noalign{\global\let\adl@draw\@dashdrawstore
           \vskip\belowrulesep}}
\makeatother

\usepackage{xcolor}
\hypersetup{
    colorlinks=true,
    citecolor=black,
    linkcolor=black,
    urlcolor=blue,
}

\usepackage{soul}
\setuldepth{Mask-CTC}

\title{A Comparative Study on Non-Autoregressive Modelings for\\Speech-to-Text Generation}
\name{
    \begin{tabular}{c}
    Yosuke Higuchi$^1$, 
    Nanxin Chen$^2$, 
    Yuya Fujita$^3$, 
    Hirofumi Inaguma$^4$, 
    Tatsuya Komatsu$^5$,\\
    Jaesong Lee$^6$, 
    Jumon Nozaki$^{4,5}$, 
    Tianzi Wang$^2$, 
    Shinji Watanabe$^7$
    \end{tabular}
}
\address{
    $^1$Waseda University, 
    $^2$Johns Hopkins University, 
    $^3$Yahoo Japan Corporation, 
    $^4$Kyoto University,\\
    $^5$LINE Corporation, 
    $^6$Naver Corporation, 
    $^7$Carnegie Mellon University
}
\begin{document}
\ninept

\maketitle
\setlength{\abovedisplayskip}{4pt}
\setlength{\belowdisplayskip}{4pt}

\begin{abstract}
Non-autoregressive (NAR) models simultaneously generate multiple outputs in a sequence, 
which significantly reduces the inference speed at the cost of accuracy drop compared to autoregressive baselines. 
Showing great potential for real-time applications, 
an increasing number of NAR models have been explored in different fields to mitigate the performance gap against AR models.
In this work, we conduct a comparative study of various NAR modeling methods for end-to-end automatic speech recognition (ASR). 
Experiments are performed in the state-of-the-art setting using ESPnet. 
The results on various tasks provide interesting findings for developing an understanding of NAR ASR, 
such as the accuracy-speed trade-off and 
robustness against long-form utterances. 
We also show that the techniques can be combined for further improvement and 
applied to NAR end-to-end speech translation. 
All the implementations are publicly available to encourage further research in NAR speech processing.
\end{abstract}

\begin{keywords}
Non-autoregressive sequence generation, end-to-end speech recognition, end-to-end speech translation
\end{keywords}

\section{Introduction}
\label{sec:intro}
In the last decade, 
deep learning has brought remarkable success in automatic speech recognition (ASR)~\cite{hinton2012deep, graves2013speech}, 
which has become a central user interface in various IoT applications. 
Much of the recent research progress is attributed to improvement in the end-to-end system~\cite{graves2014towards, chorowski2015attention, chan2016listen}, 
where an ASR model is trained to directly optimize speech-to-text conversion. 
Owing to the well-established sequence-to-sequence modeling techniques~\cite{graves2012sequence, sutskever2014sequence, bahdanau2014neural} 
and more sophisticated neural network architectures~\cite{dong2018speech, kriman2020quartznet, gulati2020conformer}, 
end-to-end ASR systems have achieved comparable results with those of the conventional hybrid systems~\cite{chiu2018state, luscher2019rwth, karita2019comparative}. 

\begin{figure}[t]
    \centering
    \vspace{2mm}
    \includegraphics[width=0.89\columnwidth]{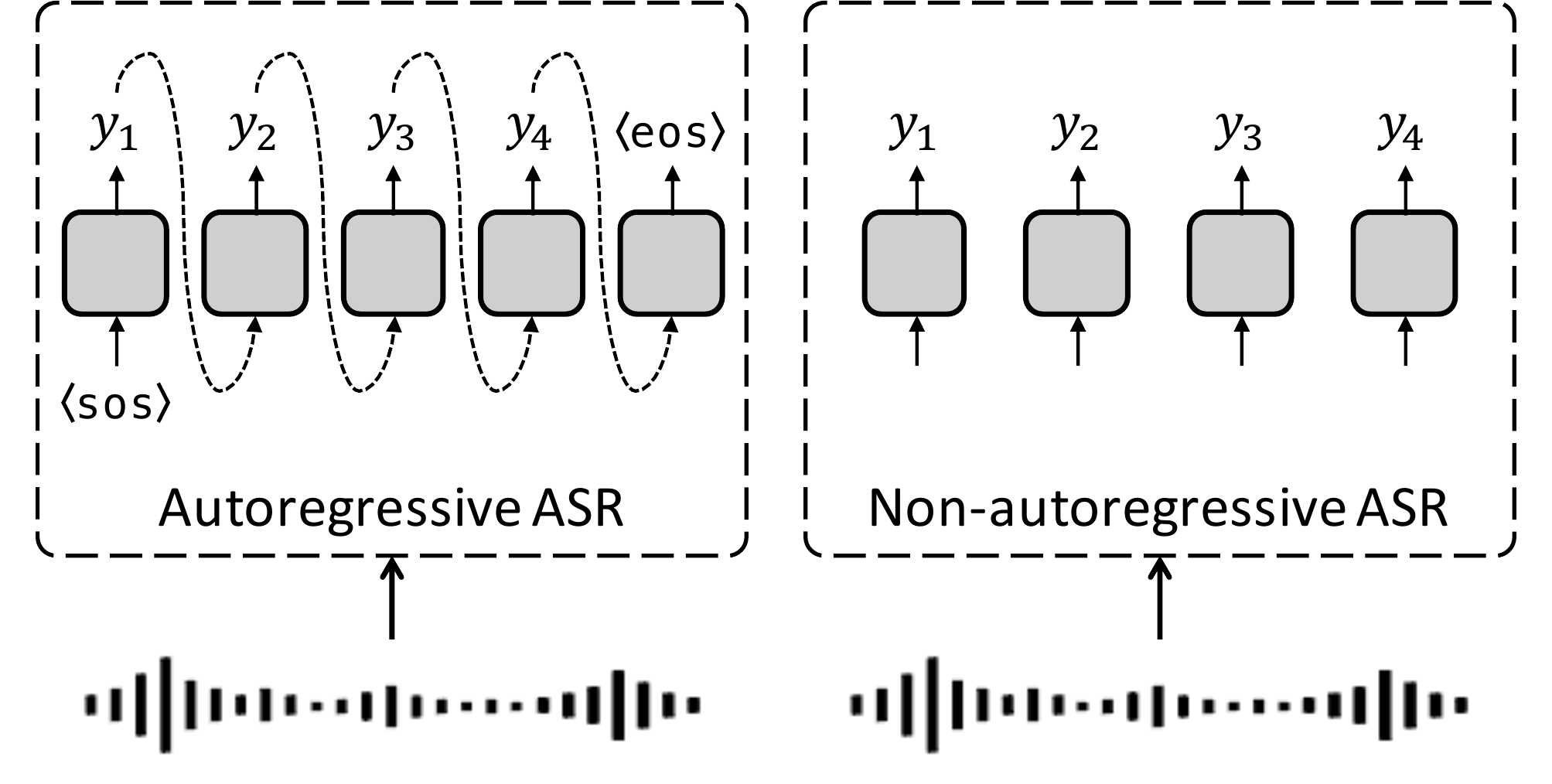}
    \vspace{-4.2mm}
    \caption{Illustrations of autoregressive and non-autoregressive ASR.}
    \label{fig:ar_nar}
    \vspace{-3mm}
\end{figure}

Current state-of-the-art end-to-end ASR systems are based on \textit{autoregressive} (AR) models~\cite{graves2012sequence, bahdanau2014neural}, 
where each token prediction is conditioned on the previously generated tokens (Figure~\ref{fig:ar_nar} left). 
Such a generation process can lead to slow inference speed, 
requiring $L$-step incremental calculations to generate an $L$-length sequence. 
\textit{Non-autoregressive} (NAR) models~\cite{graves2006connectionist, gu2017non}, 
in contrast, 
permit generating multiple tokens in parallel, 
which significantly speeds up the decoding process (Figure~\ref{fig:ar_nar} right). 
However, such simultaneous predictions often prevent the NAR models from learning the conditional dependencies between output tokens, 
worsening the recognition accuracy compared to AR models. 

Fast inference is one of the crucial factors for deploying deep learning models to real-world applications. 
ASR systems, in particular, are desired to be fast and light in numerous scenes, 
such as in spoken dialogue systems, where users prefer quick interactions with a conversational agent. 
Accordingly, various attempts have been actively made to develop and improve NAR end-to-end ASR models~\cite{chen2020non, chan2020imputer, higuchi2020mask, bai2020listen, fujita2020insertion, tian2020spike, chi2020align}, 
inspired by the great success of NAR modeling techniques in neural machine translation~\cite{libovicky2018end, lee2018deterministic, stern2019insertion, gu2019levenshtein, ghazvininejad2019mask, saharia2020non, ma2019flowseq}. 
However, 
while an increasing number of NAR models have been proposed and shown their effectiveness,  
the research community lacks a comprehensive study for comparing different NAR methods in a fair experimental setting. 
Hence, it remains unclear what the advantages and disadvantages each model has when compared to other models. 

Our work aims to conduct a comparative study on NAR modeling methods for end-to-end ASR. 
We have made an effort to cover a wide variety of methods, 
including 
a standard connectionist temporal classification (CTC)-based model~\cite{graves2014towards}; 
Mask-CTC~\cite{higuchi2020mask} and Improved Mask-CTC~\cite{higuchi2021improved} based on masked language modeling~\cite{devlin2019bert, ghazvininejad2019mask}; 
Align-Denoise~\cite{chen2021align} based on refinement training~\cite{lee2018deterministic}; 
Insertion Transformer and KERMIT~\cite{fujita2020insertion} based on insertion-based modeling~\cite{stern2019insertion, chan2019kermit}; 
intermediate CTC~\cite{lee2021intermediate} and self-conditioned CTC~\cite{nozaki2021relaxing} based on regularization techniques; and 
a continuous integrate-and-fire (CIF)-based NAR model (CIF-NA)~\cite{dong2020cif}. 
All the models are fairly evaluated in the state-of-the-art setup using ESPnet~\cite{watanabe2018espnet}, 
adopting Conformer~\cite{gulati2020conformer} for the network architecture. 

The contributions of this work are summarized as follows:
\begin{itemize}
    \item We conduct comparative experiments on a variety of NAR ASR models using various ASR tasks. 
    In addition to comparing the accuracy and speed, 
    we further analyze the results to help develop a deep understanding of NAR ASR. 
    \item We show that different NAR techniques can be combined to improve the performance and 
    applied to other NAR speech-to-text tasks, e.g., end-to-end speech translation. 
    \item We provide reproducible implementations and recipes used in our experiments, hoping to encourage further research in NAR speech processing. 
\end{itemize}

\section{Related works}
\label{sec:related_works}

\begin{table}[t]
    \centering
    \caption{Comparison of various NAR end-to-end ASR models.}
    \label{tb:nar_models}
    \renewcommand{\arraystretch}{0.93}
    \scalebox{0.91}{
     \begin{tabular}{lcccc}
        \toprule
        \textbf{Model} &\textbf{\#iter} & \textbf{Processing unit} & \textbf{CTC} \\
        \midrule
        A-CMLM~\cite{chen2020non} & 3 & token & \\
        Imputer~\cite{chan2020imputer} & 8 & frame & \checkmark \\
        LASO~\cite{bai2020listen} & 1 & token &  \\
        Spike-Triggered~\cite{tian2020spike} & 1 & token & \checkmark \\
        Mask-CTC~\cite{higuchi2020mask} & 10 & token & \checkmark \\
        Improved Mask-CTC~\cite{higuchi2021improved} & 5 & token & \checkmark \\
        Align-Refine~\cite{chi2020align} & 5 & frame & \checkmark \\
        Align-Denoise~\cite{chen2021align} & 1 & frame & \checkmark \\
        Insertion Transformer~\cite{fujita2020insertion} & $\simeq \log_2(L)$ & token \\
        KERMIT~\cite{fujita2020insertion} & $\simeq \log_2(L)$ & token & \checkmark \\
        Intermediate CTC~\cite{lee2021intermediate} & 1 & frame & \checkmark \\
        Self-conditioned CTC~\cite{nozaki2021relaxing} & 1 & frame & \checkmark \\
        CIF-NA~\cite{yu2021boundary} & 1 & token & \checkmark \\
        \bottomrule
     \end{tabular}}
     \vspace{-3mm}
\end{table}

In Table~\ref{tb:nar_models}, 
we list several NAR end-to-end ASR models to help compare and understand them at a glance. 
Here, we include some important aspects for the comparison. 
The number of decoding iterations (\textbf{\#iter}) can be increased to improve an output sequence 
at the expense of extra computations, 
which enables a model to generate more valid tokens conditioned on previously generated tokens in a semi-autoregressive manner~\cite{lee2018deterministic,ghazvininejad2019mask}. 
\textbf{Processing unit} is a unique property in NAR ASR, which can be either frame-level or token-level. 
Token-level processing reduces the speed and computational cost during inference, 
which is especially important when a model performs the iterative prediction. 
However, it needs additional efforts to estimate or adjust the length of an output sequence~\cite{gu2017non}. 
Frame-level processing, on the other hand, is prone to slow inference with the requirement of computational resources, 
but it does not require the length prediction. 
\textbf{CTC} indicates the usage of connectionist temporal classification (CTC)~\cite{graves2006connectionist}, which is a promising NAR modeling method for ASR. 

CTC is the very fundamental method for realizing NAR end-to-end ASR. 
CTC makes a strong conditional independence assumption between token frame predictions, 
enabling the model to perform fast inference while limiting the recognition accuracy compared to other end-to-end ASR models~\cite{battenberg2017exploring}. 
Inspired by the conditional masked language model (CMLM)~\cite{ghazvininejad2019mask}, 
Audio-CMLM (A-CMLM)~\cite{chen2020non} effectively learns the conditional distribution of output tokens over a partially observed sequence through the NAR mask prediction task~\cite{devlin2019bert}. 
Imputer~\cite{chan2020imputer} and Mask-CTC~\cite{higuchi2020mask, higuchi2021improved} combine CTC with CMLM to improve frame-level or token-level CTC predictions, 
getting rid of the cumbersome length prediction required in the previous approach. 
While Imputer and Mask-CTC suffer from the mismatch between training and testing conditions, 
Align-Refine~\cite{chi2020align} and Align-Denoise~\cite{chen2021align} introduce iterative refinement~\cite{lee2018deterministic} to optimize the refinement process of CTC predictions directly. 

Some of the recent efforts in NAR end-to-end ASR focus on improving the performance of the standard CTC-based model itself. 
Intermediate CTC~\cite{lee2021intermediate} and self-conditioned CTC~\cite{nozaki2021relaxing} 
apply auxiliary CTC losses to intermediate layers as in~\cite{tjandra2020deja}, 
which effectively enhances the intermediate representations and leads to improved CTC performance. 
Convolution-based neural network architectures have been shown to improve the CTC-based and the other end-to-end ASR models in general~\cite{higuchi2021improved, ng2021pushing, majumdar2021citrinet}. 
When a large amount of speech data is available for pre-training, 
powerful speech representations learned by wav2vec 2.0~\cite{baevski2020wav2vec} can significantly boost the performance of CTC~\cite{ng2021pushing}. 

Another direction for NAR ASR is based on insertion-based modeling, 
which permits the model for generating tokens in an arbitrary order without the left-to-right constraint in AR models. 
Showing promising results in neural machine translation, 
Insertion Transformer~\cite{stern2019insertion} and Kontextuell Encoder Representations Made by Insertion Transformations (KERMIT)~\cite{chan2019kermit} are successfully adopted for end-to-end ASR~\cite{fujita2020insertion}.

\section{Non-autoregressive ASR}
\label{sec:nar_asr}
This section reviews NAR modeling methods for end-to-end ASR compared in our study, 
including CTC, 
Mask-CTC, Improved Mask-CTC, 
Align-Denoise, 
Insertion Transformer, KERMIT, 
intermediate CTC, self-conditioned CTC, and 
CIF-NA. 
We have made an effort to cover a wide variety of methods, 
each of which has a unique capability as an NAR model, 
as described in Section~\ref{sec:related_works}. 

\noindent\textbf{Notations:} We formulate end-to-end ASR as a sequence mapping between a $T$-length input sequence $X \!=\! (\bm{\mathrm{x}}_t \in \mathbb{R}^D| t\!=\!1,\dots,T)$ and an $L$-length output sequence $Y \!=\! ( y_l \in \mathcal{V} | l\!=\!1,\dots,L )$. 
Here, $\bm{\mathrm{x}}_t$ is a $D$-dimensional acoustic feature at frame $t$, 
$y_l$ an output token at position $l$, and $\mathcal{V}$ a vocabulary.

\subsection{Connectionist temporal classification (CTC)}

\label{sec:ctc}
CTC~\cite{graves2006connectionist} predicts a frame-level alignment sequence $Z=(z_t \in \mathcal{V} \cup \{\epsilon\}| t=1,\dots,T)$, which is obtained by introducing a special blank token $\epsilon$ into the output sequence $Y$. 
Based on the conditional independence assumption per token frame prediction, 
CTC models the conditional probability $P(Y|X)$ by marginalizing over all paths (frame alignments) as: 
\begin{equation}
    \label{eq:p_ctc}
    P_{\mathsf{ctc}} (Y | X) = \sum_{Z \in \mathcal{B}^{-1} (Y)} \prod_{t=1}^{T} P (z_t | X), 
\end{equation}
where $\mathcal{B}^{-1}(Y)$ denotes all possible paths compatible with $Y$. 
The CTC objective is defined by the negative log-likelihood of Eq.~\eqref{eq:p_ctc}: 
\begin{equation}
    \mathcal{L}_{\mathsf{ctc}} = -\log P_{\mathsf{ctc}}(Y|X). 
    \label{eq:ctc_loss}
\end{equation}
During inference, we use the best path decoding~\cite{graves2006connectionist} to generate an output sequence, 
where the most probable tokens $\argmax_Z P(Z|X)$ are selected at each frame, and an output sequence is obtained by suppressing repeated tokens and removing blank symbols.

\subsection{Mask-CTC}
Mask-CTC~\cite{higuchi2020mask} is built upon an encoder-decoder structure, 
where the CTC loss (Eq.~\eqref{eq:p_ctc}) is applied to the encoder output, and 
the decoder adopts the conditional masked language model (CMLM)~\cite{ghazvininejad2019mask, chen2020non}. 
The CMLM decoder is trained to predict output tokens $Y_{\mathsf{mask}} \in Y$, given a partially observed ground-truth sequence $Y_{\mathsf{obs}}\!=\!Y \setminus Y_{\mathsf{mask}}$: 
\begin{equation}
      P_{\mathsf{cmlm}} (Y_{\mathsf{mask}} | Y_{\mathsf{obs}}, X) = 
      \prod_{y \in Y_{\mathsf{mask}}} P (y | Y_{\mathsf{obs}}, X), 
      \label{eq:p_cmlm}
\end{equation}
where $Y_{\mathsf{mask}}$ are obtained by randomly replacing ground-truth tokens with a special mask token \texttt{<MASK>}. 
The objective of the CMLM decoder is defined as:
\begin{equation}
    \mathcal{L}_{\mathsf{cmlm}} = - \log P_{\mathsf{cmlm}} (Y_{\mathsf{mask}} | Y_{\mathsf{obs}}, X). 
    \label{eq:l_cmlm}
\end{equation}
The final loss of Mask-CTC is defined as a weighted sum of the standard CTC loss $\mathcal{L}_{\mathsf{ctc}}$ in Eq.~\eqref{eq:ctc_loss} and $\mathcal{L}_{\mathsf{cmlm}}$ in Eq.~\eqref{eq:l_cmlm}.
During inference, 
an output sequence of CTC is first obtained from the encoder, and then
the decoder refines the CTC output through the mask prediction process based on Eq.~\eqref{eq:p_cmlm}. 
By masking low-confidence tokens in the CTC output and predicting the masked tokens based on the other high-confidence unmasked tokens, 
errors from the conditional independence assumption are expected to be recovered. 

\vspace{-1mm}
\subsection{Improved Mask-CTC}
\vspace{-1mm}
One limitation of Mask-CTC is that the length of an output sequence cannot be changed from that of the CTC output during inference, 
making it difficult to recover deletion and insertion errors. 
To overcome this problem, 
Mask-CTC is enhanced by introducing a length prediction network in the CMLM decoder~\cite{higuchi2021improved}. 
The length prediction network is trained to predict the length of a partial target sequence from a masked token. 
For example, 
given a ground-truth sequence $Y\!=\![y_1, y_2, y_3, y_4]$ and 
its masked sequence $[y_1, \text{\texttt{<MASK>}}, y_4, \text{\texttt{<MASK>}}]$, 
symbols \texttt{2} and \texttt{0} are predicted from each masked position, 
indicating the length of the corresponding partial sequence in $Y$. 
With this length prediction network, during inference, 
the number of \text{\texttt{<MASK>}} at each masked position is first modified based on the predicted length. 
Each mask is then predicted conventionally as in Eq.~\eqref{eq:p_cmlm}, 
allowing the model to change the sequence length dynamically by deleting and inserting mask tokens.

\subsection{Align-Denoise}

Align-Denoise~\cite{chen2021align} is built on the prior work of Align-Refine~\cite{chi2020align} and Imputer~\cite{chan2020imputer} and adopts a similar encoder-decoder structure on frames.
CTC is applied to both the encoder output and decoder output. 
Instead of taking multiple steps to get the intermediate results as Align-Refine, Align-Denoise generates noisy CTC alignment $\tilde Z$ during the training and requires a single iteration for decoding.
The training objective is based on the CTC loss in Eq.~\eqref{eq:ctc_loss} and a ground-truth alignment $Z_\mathsf{gt} = \argmax_Z P(Z|X)$:
\begin{equation}
    \mathcal{L}_\mathsf{aligndenoise} = \mathbb{E}_{\tilde Z \sim q(\tilde Z \mid X, Z_\mathsf{gt})}
    \left[\log P_{\mathsf{ctc}}(Y|X,\tilde Z) \right], 
\end{equation}
where $q(\cdot)$ is the noisy function for generating the noisy alignment.
The idea is similar to the denoising autoencoder~\cite{vincent2008extracting} (DAE), 
where the input to the decoder is the ground truth alignment with a certain level of noise.

\subsection{Intermediate CTC}

Intermediate CTC~\cite{lee2021intermediate} extends the CTC-based modeling with a regularization loss.
During training, a sequence of intermediate representations $X_\mathsf{inter}$ is obtained from an intermediate layer of the encoder, and its intermediate CTC loss is computed as:
\begin{align}
\label{eq:inter_loss}
    \mathcal{L}_\mathsf{inter} = - \log P_{\mathsf{ctc}}(Y|X_\mathsf{inter}). 
\end{align} 
The final loss is a weighted sum of the original CTC loss in Eq.~\eqref{eq:ctc_loss} and intermediate loss.
During inference, the intermediate loss is unused, and the model is treated as an ordinary CTC model.

\subsection{Self-conditioned CTC}

Self-conditioned CTC~\cite{nozaki2021relaxing} extends intermediate CTC by exploiting the intermediate representations $X_{\mathsf{inter}}$ for conditioning the subsequent encoder layers. 
During both training and inference, each token posterior distribution in the intermediate layers $A_{\mathsf{inter}}=\mathrm{softmax}(X_{\mathsf{inter}})$ 
is fed back to the input of the next layer, 
making the subsequent encoder layers conditioned on the intermediate predictions.
The self-conditioned CTC loss is defined as:
\begin{align}
     \mathcal{L}_\mathsf{selfcond} = - \log P_{\mathsf{ctc}}(Y|X, A_{\mathsf{inter}}).
\end{align}
The final loss is a weighted sum of the intermediate CTC loss in Eq.~\eqref{eq:inter_loss} and the self-conditioned CTC loss.

\subsection{Insertion Transformer}

In the case of insertion-based NAR models, 
the conditional probability $P(Y|X)$ is modeled by marginalizing over  insertion order $\pi$:
\begin{align}
\label{eq:ins_model}
    P_{\mathsf{ins}}(Y|X) & = \sum _{\pi} P(Y, {\pi}|X) = \sum _{\pi} P(Y^{\pi} |X) P({\pi}).
\end{align}
Insertion order ${\pi}$ represents the permutation of tokens in a sequence $Y$, 
e.g., if $L=4$ and ${\pi}=(3,1,2,4)$, $Y^{\pi} = (y_3, y_1, y_2, y_4)$.

During training, an upper bound of negative log-likelihood is minimized under a prior distribution over insertion order $\pi$:
\begin{align}
    \mathcal{L}_{\mathsf{ins}} =  - \sum _{\pi} P(\pi) \log P(Y^{\pi}|X) \geq - \log P_{\mathsf{ins}}(Y|X).
\end{align}
Insertion Transformer~\cite{stern2019insertion} trains Transformer, with an encoder-decoder structure, 
to predict a token and its position to be inserted, 
which aims to model $P(Y^{\pi}|X)$ in Eq.~\eqref{eq:ins_model}. 
When $P({\pi})$ is defined by the balanced binary tree (BBT)-based insertion order~\cite{stern2019insertion}, decoding finishes empirically with $\log_2(L)$ iterations. 
The BBT order is to insert the centermost tokens of the current hypothesis. 
For example, given an output sequence with $L=7$, the hypothesis grows based on the tree structure like $(y_4) \rightarrow (y_2, y_4, y_6) \rightarrow (y_1, y_2, y_3, y_4, y_5, y_6, y_7)$.

\subsection{KERMIT}

KERMIT~\cite{chan2019kermit} is a variant of Insertion Transformer. 
Its basic formulation is the same but only the Transformer encoder is used to predict a token and its position to be inserted.

KERMIT can be trained with the CTC loss in a multi-task learning manner~\cite{fujita2020insertion},
which makes the CTC alignment prediction conditioned on a partial hypothesis from the insertion-based decoding.
Given the posterior probability of a CTC alignment at $k$-th decoding step as $P(Z|Y^{{\pi}^{k}}, X)$, 
the objective of KERMIT is to minimize the following negative log-likelihood:
\begin{align}
    \label{eq:kermit_model}
    \mathcal{L}_{\mathsf{kermit}} = \mathcal{L}_{\mathsf{ins}} - \lambda_{\mathsf{kermit}}  \log \sum_{Z \in \mathcal{B}^{-1} (Y)} P (Z | Y^{{\pi}^{k}}, X),
\end{align}
where $\lambda_{\mathsf{kermit}}$ is a tunable weight on the CTC loss. 
During inference, the CTC decoding is performed using $P (Z | Y^{{\pi}^{k}}, X)$ in Eq.~\eqref{eq:kermit_model}, and 
it can be terminated at any number of iterations.

\subsection{CIF-NA}

Continuous integrate-and-fire (CIF)\cite{dong2020cif} provides a soft and monotonic alignment in the encoder-decoder framework. 
CIF first learns information weights $(\alpha_{1},...\alpha_{T})$ from the encoder output $X_\mathsf{enc}$ and accumulates the weights from left to right to locate the acoustic boundary. 
Then, the acoustic embedding $C=(c_1, .., c_L)$ for each target token is obtained by integrating the encoder sates based on their estimated weights.
In this paper, we investigated CIF with an NAR decoder, noted as CIF-NA. 
Following the prior work~\cite{yu2021boundary}, 
both the encoder states and acoustic embeddings are fed into the decoder to predict the probability of output tokens in parallel:
\begin{align}
    \mathcal{L}_\mathsf{cif}=-\text{log}\prod_{l=1}^{L} P(y_l|C, X_\mathsf{enc}).
\end{align}
CIF also adopts a quantity loss to supervise the model to predict the quantity of the integrated embeddings closer to the length of a target sequence, defined as $\mathcal{L}_\mathsf{qua}=|\sum_{t=1}^T \alpha_t - L|$. 
The final loss of CIF-NA is a weighted sum of the CTC loss in Eq.~\eqref{eq:ctc_loss}, $\mathcal{L}_\mathsf{cif}$ and $\mathcal{L}_\mathsf{qua}$.

\begin{table*}[t]
    \centering
    \caption{Word error rate (WER) or character error rate (CER) on LibriSpeech-100h (LS-100), TEDLIUM2 (TED2), and CSJ-APS. \#iter denotes the number of iterations required to generate each token in an output sequence. Real time factor (RTF) was used to measure the inference speed and was evaluated on the LS-100 ``test-other'' set using CPU.}
    \label{tb:main}
    \renewcommand{\arraystretch}{0.95}
    \newcolumntype{C}{>{\centering\arraybackslash}p{7mm}}
    \scalebox{0.93}{
    \begin{tabular}{p{6mm}lcccCCCCCCCCC}
        \toprule
        \multicolumn{2}{l}{\multirow{3}{*}[-3pt]{\textbf{Model}}} & \multirow{3}{*}[-3pt]{\textbf{\#iter}} &  \multicolumn{2}{c}{\textbf{Inference speed}} & \multicolumn{4}{c}{\textbf{LS-100} (WER)} & \multicolumn{2}{c}{\textbf{TED2} (WER)} &  \multicolumn{3}{c}{\textbf{CSJ-APS} (CER)} \\
        \cmidrule(l{0.2em}r{0.4em}){4-5}
        \cmidrule(l{0.4em}r{0.4em}){6-9}
        \cmidrule(l{0.4em}r{0.4em}){10-11}
        \cmidrule(l{0.4em}r{0.4em}){12-14}
        & & & \multirow{2}{*}[0pt]{RTF} & \multirow{2}{*}[0pt]{Speedup} & \multicolumn{2}{c}{dev} & \multicolumn{2}{c}{test} & \multirow{2}{*}[0pt]{dev} & \multirow{2}{*}[0pt]{test} & \multirow{2}{*}[0pt]{eval1} & \multirow{2}{*}[0pt]{eval2} & \multirow{2}{*}[0pt]{eval3} \\
        & & & & & clean & other & clean & other & & & & & \\
        \midrule
        \multirow{4}{*}[0pt]{AR} & CTC/attention & $L$ & 0.341 & 1.00$\times$  & \phantom{0}6.8 & 18.8 & \phantom{0}7.4 & 19.0 & 11.6 & 8.7 & 5.4 & 4.0 & \phantom{0}9.8 \\
        & \hspace{1mm} + beam-search & $>L$ & 3.419 & 0.10$\times$ & \textbf{\phantom{0}6.3} & \textbf{18.2} & \textbf{\phantom{0}6.8} & \textbf{18.5} & 10.4 & 8.4 & \textbf{5.1} & \textbf{3.8} & \textbf{\phantom{0}9.0} \\
        & Transducer & $L$ & 0.069 & 4.94$\times$ & \phantom{0}7.3 & 19.9 & \phantom{0}7.6 & 19.9 & \phantom{0}9.6 & 9.2 & 6.3 & 4.5 & 10.6 \\
        & \hspace{1mm} + beam-search & $>L$ & 0.234 & 1.46$\times$ & \phantom{0}6.4 & 18.8 & \textbf{\phantom{0}6.8} & 18.9 & \textbf{\phantom{0}8.6} & \textbf{8.2} & 5.2 & 4.1 & 10.0 \\
        \cdashlinelr{1-14}
        \multirow{9}{*}[0pt]{NAR} & CTC & 1 & 0.059 & 5.78$\times$ & \phantom{0}7.4 & 20.5 & \phantom{0}7.8 & 20.8 & \phantom{0}8.9 & 8.6 & 5.4 & 4.0 & \phantom{0}9.6 \\
        & Mask-CTC & 10 & 0.063 & 5.41$\times$ & \phantom{0}7.2 & 20.3 & \phantom{0}7.5 & 20.6 & \phantom{0}8.9 & 8.5 & 5.6 & 4.0 & \phantom{0}9.6\\
        & Improved Mask-CTC & 5 & 0.072 & 4.74$\times$ & \phantom{0}7.0 & 19.8 & \phantom{0}7.3 & 20.2 & \phantom{0}8.8 & 8.3 & 5.5 & 4.0 & \phantom{0}9.5 \\
        & Align-Denoise & 1 & 0.073 & 4.67$\times$ & \phantom{0}8.0 & 22.3 & \phantom{0}8.4 & 22.5 & \phantom{0}9.0 & 8.7 & 5.4 & \textbf{3.7} & \textbf{\phantom{0}9.1}\\
        & Intermediate CTC & 1 & 0.059 & 5.78$\times$ & \phantom{0}6.9 & 19.7 & \phantom{0}7.1 & 20.2 & \textbf{\phantom{0}8.5} & 8.3 & 5.6 & 4.1 & \phantom{0}9.8\\
        & Self-conditioned CTC & 1 & 0.059 & 5.78$\times$ & \textbf{\phantom{0}6.6} & \textbf{19.4} & \textbf{\phantom{0}6.9} & \textbf{19.7} & \phantom{0}8.7 & \textbf{8.0} & \textbf{5.3} & \textbf{3.7} & \textbf{\phantom{0}9.1} \\
        & KERMIT & $\simeq \log_2(L)$ & 0.361 & 1.06$\times$ & \phantom{0}7.1 & 19.7 & \phantom{0}7.4 & 20.2 & \phantom{0}9.1 & 8.2 & 5.4 & \textbf{3.7} & \phantom{0}9.5\\
        & Insertion Transformer & $\simeq \log_2(L)$ & 0.083 & 4.11$\times$ & 16.0 & 27.3 & 16.2 & 27.4 & -- & -- & -- & -- & -- \\
        & CIF-NA$^\dagger$ & 1 & 0.073 & 4.67$\times$ & 15.4 & 34.0 & 15.7 & 34.6 & -- & -- & -- & -- & -- \\
        \bottomrule
     \end{tabular}}\\
     {\footnotesize $^\dagger$According to discussions with the author of CIF~\cite{dong2020cif}, CIF-NA suffers from the degradation due to the difficulty of acoustic boundary decisions. }
     \vspace{-2mm}
\end{table*}
\section{Experiments}
Aiming to compare the NAR models in Section~\ref{sec:nar_asr}, 
we conducted ASR experiments using ESPnet~\cite{watanabe2018espnet,watanabe2020espnet}. 
In addition to the NAR models,
we evaluated autoregressive (AR) models,
including the attention-based sequence-to-sequence model with the CTC/attention objectives~\cite{watanabe2017hybrid, karita2019improving} and
Conformer-Transducer~\cite{guo2021recent}.
The recognition accuracy was evaluated based on word error rate (WER) or character error rate (CER), depending on a task, and 
the inference speed was measured using real time factor (RTF). 

\subsection{Experimental setup}
\noindent{\textbf{Data:}}
The main experiments were carried out using three datasets, including LibriSpeech (LS)~\cite{panayotov2015librispeech}, TEDLIUM2 (TED2)~\cite{rousseau2014enhancing}, and Corpus of Spontaneous Japanese (CSJ)~\cite{maekawa2000spontaneous}. 
LS consists of utterances from read English audiobooks, and 
we used the 100-hour subset (LS-100) for training. 
TED2 contains utterances from English Ted Talks, and we used the 210-hour training data. 
CSJ includes Japanese public speeches on different academic topics, and 
we used the 271-hour subset of academic presentation speech (CSJ-APS) for training. 
For LS-100 and TED2, 
we used the standard validation and test sets for tuning hyper-parameters and evaluating performance, respectively. 
Specifically, for LS-100, 
the validation and test sets are divided into ``clean'' and ``other'' based on the quality of the recorded utterances. 
For CSJ-APS, 
we used a part of the training set as a validation set and 
the official evaluation sets (``eval1'', ``eval2'', and ``eval3'') for testing. 
For LS-100 and TED2, 
we used 300 to 500 subwords for tokenizing output texts, 
which were constructed from each training set using SentencePiece~\cite{kudo2018subword}. 
For CSJ-APS, 
we used Japanese syllable characters (Kana) and Chinese characters (Kanji), 
which resulted in 2753 distinct tokens. 

We also evaluated several models under noisy conditions using CHiME-4~\cite{vincent2017analysis}. 
CHiME-4 contains English recordings in everyday noisy environments, 
including a cafe, a street junction, public transport, and a pedestrian area. 
We combined the CHiME-4 and Wall Street Journal (WSJ)~\cite{paul1992design} datasets to obtain 
a 190-hour training set. 
We used the validation and test sets provided by CHiME4 for tuning hyper-parameters and evaluating performance, respectively. 
Characters (Latin alphabets) were used for tokenizing target texts. 

As input speech features, 
we extracted 80 mel-scale filterbank coefficients with three-dimensional pitch features using Kaldi~\cite{povey2011kaldi}. 
To avoid overfitting, 
we applied speed perturbation~\cite{ko2015audio} and SpecAugment~\cite{park2019specaugment} to the input speech from LS-100, TED2, and CSJ-APS, and 
only SpecAugment to the input speech from CHiME4. 

\noindent{\textbf{Network architecture:}}
All the end-to-end ASR models were constructed based on 
the Conformer-based architecture~\cite{gulati2020conformer, guo2021recent}. 
For the self-attention module, 
the number of heads $d_{\mathrm h}$, 
the dimension of a self-attention layer $d_{\mathrm{model}}$, and 
the dimension of a feed-forward network $d_{\mathrm{ff}}$
were set to 4, 256, and 1024, respectively. 
The kernel size of the convolution module was set to 15. 
For the models with the encoder-decoder structure (i.e., CTC/attention, Mask-CTC, Improved Mask-CTC, Align-Denoise, Insertion Transformer, and CIF-NA), 
the encoder consisted of two convolutional neural network (CNN)-based downsampling layers followed by a stack of 12 Conformer encoder blocks. 
The decoder consisted of 6 Transformer decoder blocks, 
where the self-attention module had the same configuration as the Conformer block, 
except $d_{\mathrm{ff}}$ was increased to 2048 to match the number of parameters in the Conformer block. 
For the models without the decoder (i.e., CTC, KERMIT, Intermediate CTC, Self-conditioned CTC), 
we used two CNN-based downsampling layers followed by a stack of 18 Conformer encoder blocks.
Conformer-Transducer consisted of the 18-layer encoder and
a single long short-term memory (LSTM) layer for the decoder.
Note that the number of parameters in all the ASR models was around 30M. 

\noindent{\textbf{Training and decoding configurations:}}
The ASR models were trained using the Adam optimizer~\cite{kingma2015adam} with $\beta_1\!=\!0.9$, $\beta_2\!=\!0.98$, $\epsilon\!=\!10^{-9}$, and 
the Noam learning rate scheduling~\cite{vaswani2017attention}. 
Warmup steps (e.g., 4k to 25k) and a learning rate factor (e.g., 1.0 to 10.0) were tuned for each model. 
We followed the same setup as in~\cite{karita2019comparative, guo2021recent} for regularization hyperparameters (e.g., dropout rate and label-smoothing weight).
The models were trained up to 300 epochs until convergence. 
For evaluation, 
a final model was obtained by averaging model parameters over 5 to 30 checkpoints with the best validation performance. 
For decoding with the CTC-based NAR models, 
we performed the best path decoding of CTC~\cite{graves2006connectionist} 
to keep the inference process NAR. 
For evaluating the AR models,
we applied beam search decoding with beam sizes of 1 (i.e., greedy decoding) and 10. 
All of the decodings were done without using external language models (LMs). 
Decoding hyper-parameters that are unique to each model were tuned following the previous works.
To evaluate the inference speed, RTF was measured using Intel(R) Xeon(R) Silver 4114 CPU, 2.20GHz on the same machine environment. 
All of the implementations are publicly available\footnote{
\href{https://github.com/espnet/espnet}{\ul{Mask-CTC}},
\href{https://github.com/YosukeHiguchi/espnet/tree/maskctc_dlp}{\ul{Improved Mask-CTC}},
\href{https://github.com/bobchennan/espnet/tree/align_denoise}{\ul{Align-Denoise}},
\href{https://github.com/espnet/espnet}{\ul{Intermediate CTC}},
\href{https://github.com/jumon/espnet-1/tree/selfconditioned}{\ul{Self-conditioned CTC}},
\href{https://github.com/yuyfujit/espnet/tree/insertion-based-models}{\ul{Insertion Transformer}},
\href{https://github.com/yuyfujit/espnet/tree/insertion-based-models}{\ul{KERMIT}},
\href{https://github.com/Jzmo/espnet/tree/jz-cif}{\ul{CIF-NA}}
}.

\subsection{Main results}
\label{ssec:main_results}
Table~\ref{tb:main} shows the results on LS-100, TED2, and CSJ-APS. 

\noindent\textbf{LibriSpeech-100h (LS-100):}
By comparing the results obtained from the NAR models on LS-100, 
all of our NAR models, except Insertion Transformer and CIF-NA, 
outperformed the standard CTC-based model. 
Moreover, 
on the ``clean'' sets, 
Improved Mask-CTC, intermediate CTC, self-conditioned CTC, and KERMIT achieved 
comparable performances with those of the AR models. 
On the ``other'' sets, in contrast, 
the NAR models resulted in worse performances than the CTC/attention model.
Since the ``other'' sets include utterances with lower quality than the ``clean'' sets, 
a model is required to capture dependencies between output tokens to compensate for information loss in the low-quality speech. 
With the language model mechanism included in the network structure, 
the CTC/attention model resulted in better performances on ``other'' sets, 
effectively modeling dependencies between output tokens. 

\noindent\textbf{TEDLIUM2 (TED2):} On TED2, 
all the NAR models, including the standard CTC-based model, 
achieved results competitive to those obtained from the AR models. 
Especially on the development set, 
the NAR models outperformed the CTC/attention model by a large margin. 
We further discuss this interesting outcome in Section~\ref{ssc:length_robustness}. 

\noindent\textbf{CSJ-APS:} 
All the NAR models achieved comparable performance with the AR models on CSJ-APS, 
indicating the effectiveness independent of the language. 
Especially, the performances of Align-Denoise and Self-conditioned CTC aligned with the searched results obtained from the CTC/attention model.

\noindent\textbf{Discussions on Insertion Transformer and CIF-NA:}
Insertion Transformer resulted in the performance drop because of its two unique characteristics. 
The first is that it is difficult to judge whether more tokens to be inserted or not. It is controlled by a special token that represents ``no more token to be inserted''. However, partial hypothesis including recognition errors leads to misjudging of token insertion. 
The second is that once a hypothesis encounters recognition error, it can not be recovered during inference. The error is propagated to later iterations and leads to more errors.
CIF is a novel alignment mechanism that can be adopted for various ASR scenarios. 
However, different from the original AR version, 
CIF-NA resulted in the performance degradation due to the difficulty of estimating acoustic boundaries in the English speech dataset.
According to \cite{dong2020cif,yu2021boundary}, it could achieve competitive performance in monosyllable languages with clear acoustic boundaries (e.g., Chinese Mandarin).

From the above results, 
it can be concluded that CTC is the key technique to realizing effective NAR end-to-end ASR, 
as all the well-performing models are based on CTC. 
Overall, 
self-conditioned CTC resulted in the best performance among the NAR models, 
filling the gap against the AR performance. 

\subsection{Recognition accuracy vs. inference speed}
We study the trade-off between recognition accuracy and inference speed. 
According to RTF results in Table~\ref{tb:main}, which are measured on the ``test-other'' set of LS-100, 
all the NAR models except KERMIT achieved fast inference speed compared to the AR models.
This is because of the non-autoregressive nature of NAR models, i.e., it can generate multiple tokens at a single iteration step.
The RTF improvement is achieved at the expense of quality-drop in WER. 
Especially on the ``other'' sets of LS-100, 
NAR models did not reach the performance obtained from the CTC/attention model.
In contrast, on the ``clean'' sets, 
several NAR models (e.g., self-conditioned CTC) achieved better WERs than the greedy results of the CTC/attention model. 
Improving the recognition of difficult utterances is an important problem for NAR models to be solved.

\subsection{Error analysis}
\label{ssec:error_analysis}
\begin{table}[t]
    \centering
    \caption{Error analysis on LS-100. WERs on Table~\ref{tb:main} are split into substition (sub), deletion (del), and insertion (ins) error rates.}
    \label{tb:wer_analysis}
    \renewcommand{\arraystretch}{0.92}
    \scalebox{0.95}{
    \begin{tabular}{p{0.3mm}lcccccc}
        \toprule
        \multicolumn{2}{l}{\multirow{2}{*}[-3pt]{\textbf{Model}}} & \multicolumn{3}{c}{\textbf{test-clean}} & \multicolumn{3}{c}{\textbf{test-other}} \\
        \cmidrule(l{0.4em}r{0.4em}){3-5}
        \cmidrule(l{0.4em}r{0.4em}){6-8}
        & & sub & del & ins & sub & del & ins \\
        \midrule
        \multirow{2}{*}[-1pt]{\begin{rotatebox}{90}{AR}\end{rotatebox}} & CTC/attention & 5.3 & 1.2 & 0.9 & 14.5 & 2.4 & 2.1 \\
        & \hspace{1mm} + beam-search & 5.4 & 0.6 & 0.8 & 14.7 & 1.8 & 2.1 \\
        \cdashlinelr{1-8}
        \multirow{5}{*}[-1pt]{\begin{rotatebox}{90}{NAR}\end{rotatebox}} & CTC & 6.3 & 0.7 & 0.8 & 16.6 & 2.2 & 1.9 \\
        & Improved Mask-CTC & 5.9 & 0.6 & 0.8 & 16.2 & 2.0 & 2.0 \\
        & Intermediate CTC & 5.7 & 0.6 & 0.8 & 16.2 & 1.9 & 2.1 \\
        & Self-conditioned CTC & 5.6 & 0.6 & 0.7 & 15.8 & 2.0 & 1.9 \\
        & KERMIT & 6.0 & 0.6 & 0.8 & 16.2 & 2.0 & 1.9 \\
        \bottomrule
    \end{tabular}}
    \vspace{-3mm}
\end{table}
In Table~\ref{tb:wer_analysis}, 
we break down WERs on LS-100 reported in Table~\ref{tb:main}. 
By comparing the results obtained from CTC and the other NAR models, 
it appeared that the major improvements on our NAR models are attributed to reducing substitution errors. 
Deletion and insertion error rates, in contrast, 
were kept relatively low and stayed almost the same from the CTC results, 
which can often be reduced by using wordpieces. 
The AR model handled substitution errors more effectively than the NAR models. 
However, deletion and insertion errors were higher than those of the NAR models, 
indicating that the NAR models are more effective at generating a sequence with a correct length. 
From these results, 
it can be suggested that the key for further improvement of NAR models is to mitigate substitution errors. 

\subsection{Robustness against output sequence length}
\label{ssc:length_robustness}
\begin{figure}[t]
    \begin{tabular}{c}
        \begin{minipage}[t]{\columnwidth}
            \centering
            \hspace{-5mm}
            \includegraphics[width=1.02\columnwidth]{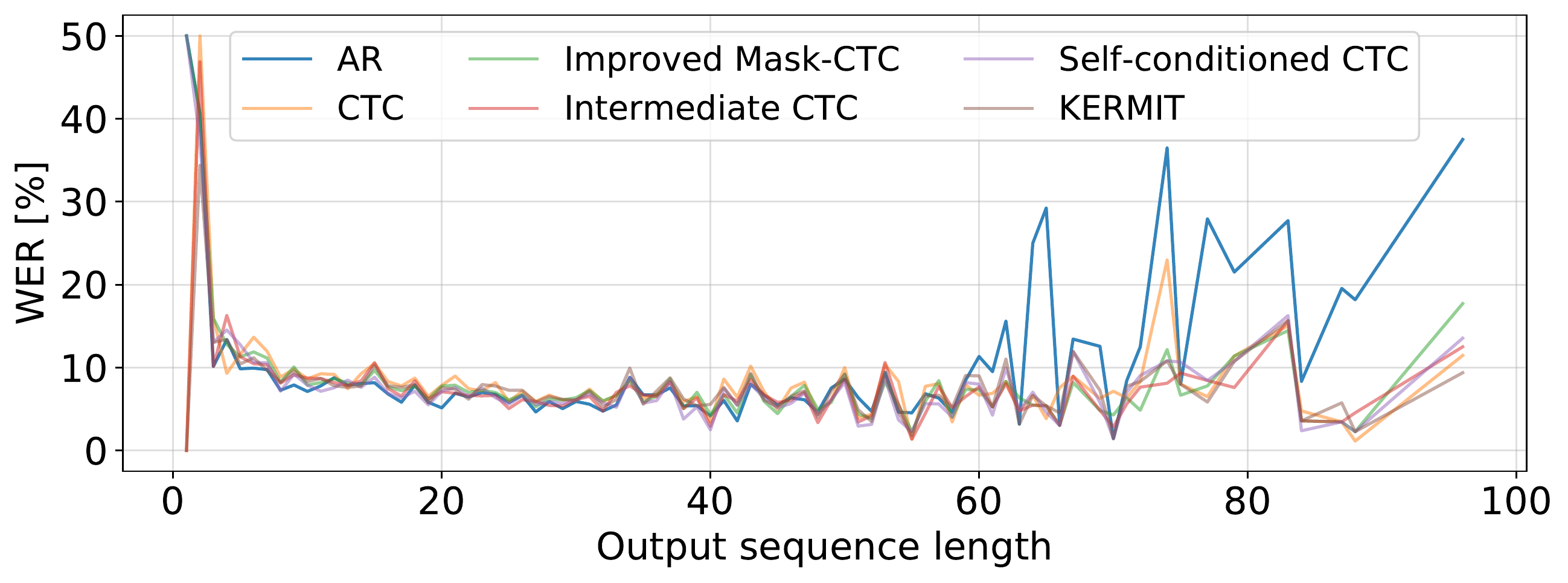}
            \end{minipage} \\
            \begin{minipage}[t]{\columnwidth}
            \centering
            \vspace{-3.5mm}
            \hspace{-5mm}
            \includegraphics[width=1.02\columnwidth]{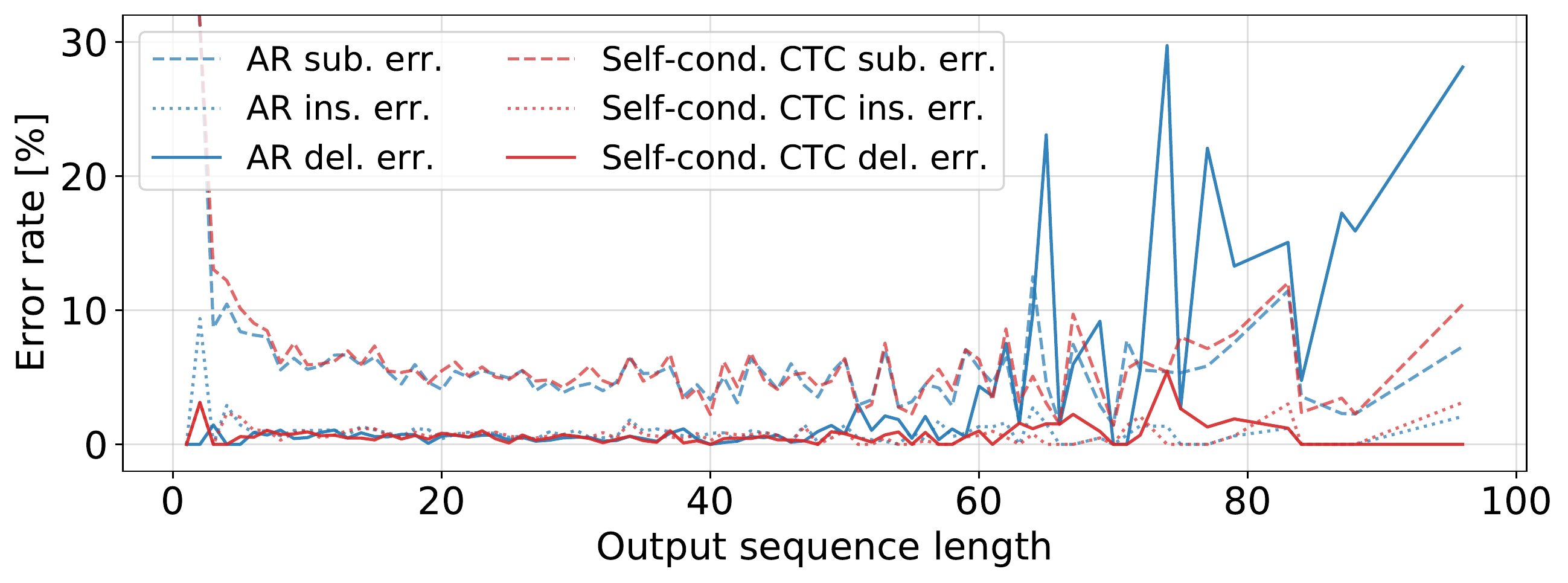}
          \end{minipage}
    \end{tabular}
    \vspace{-3mm}
    \caption{Comparison of NAR and AR (CTC/attention) models evaluated on different output sequence length. WERs are compared among the models (top), and the error components are compared between AR and Self-conditioned CTC (bottom).}
    \label{fig:length_vs_wer}
\end{figure}
Figure~\ref{fig:length_vs_wer} shows the correlation between output sequence length and error rate for the LS-100 test-clean set. 
Here, 
we observed that the performance of the AR model is prone to degradation 
when the length of an output sequence is long (Figure~\ref{fig:length_vs_wer} top). 
On the other hand, 
the NAR models successfully recognized the long sequence without having such a severe quality drop. 
To further investigate the results, 
we compared the error components between the AR and self-conditioned CTC models (Figure~\ref{fig:length_vs_wer} bottom). 
While the substitution and insertion errors were in the same range between the two models, 
the deletion error in the AR model got significantly high as the output sequence length increased, 
where we observed consecutive tokens in the sequence are completely skipped as reported in~\cite{chorowski2017towards}. 
This explains why the performance of the AR models on the TED2 development set fell behind those of the NAR models (Table~\ref{tb:main}), as the set includes long-form utterances up to 40 seconds. 

\subsection{Evaluation under noisy condition}
\begin{table}[t]
    \centering
    \vspace{-2mm}
    \caption{Word error rate on CHiME4.}
    \vspace{0.5mm}
    \renewcommand{\arraystretch}{0.95}
    \label{tb:chime4}
    \scalebox{0.95}{
    \begin{tabular}{p{0.3mm}lcccc}
        \toprule
        \multicolumn{2}{l}{\multirow{2}{*}[-3pt]{\textbf{Model}}} & \multicolumn{2}{c}{\textbf{dt05}} & \multicolumn{2}{c}{\textbf{et05}} \\
        \cmidrule(l{0.4em}r{0.4em}){3-4}
        \cmidrule(l{0.4em}r{0.4em}){5-6}
        & & real & simu & real & simu \\
        \midrule
        \multirow{2}{*}[-1pt]{\begin{rotatebox}{90}{AR}\end{rotatebox}} & CTC/attention & 14.0 & 16.0 & 22.2 & 24.1 \\
        & \hspace{1mm} + beam-search & 13.9 & 15.7 & 22.4 & 23.7 \\
        \cdashlinelr{1-6}
        \multirow{4}{*}[-1pt]{\begin{rotatebox}{90}{NAR}\end{rotatebox}} & CTC & 16.5 & 18.1 & 25.9 & 27.0 \\
        & Improved Mask-CTC & 14.9 & 16.8 & 24.6 & 25.2 \\
        & Intermediate CTC & 15.2 & 16.8 & 24.2 & 25.2 \\
        & Self-conditioned CTC & \textbf{14.6} & \textbf{16.6} & \textbf{23.7} & \textbf{24.3} \\
        \bottomrule
    \end{tabular}}
    \vspace{-2mm}
\end{table}
Table~\ref{tb:chime4} shows results under noisy conditions based on the CHiME4 task. 
Note that we only focused on the models achieving promising results on the main tasks. 
The evaluations were performed on the 1-channel track, 
where the development (dt05) and test (et05) sets included real and simulated (simu) utterances recorded from one of a single microphones on the tablet device~\cite{vincent2017analysis}. 
Comparing the results obtained from the NAR models, 
all of our NAR models outperformed the standard CTC-based model. 
However, the results were not comparable with those obtained from the AR model. 
With the nonstationary noises included in the input speech, 
it is crucial for an ASR model to attend to dependency among output tokens for generating an accurate sequence. 
As observed in the LS-100 task in Section~\ref{ssec:main_results}, 
the results suggest that the NAR models are likely to depend more on acoustic information, 
having difficulty capturing the token dependencies when the quality of an input speech is low.

\subsection{Combination of different techniques}
\begin{table}[t]
    \centering
    \caption{Word error rate (WER) on LS-100 and CHiME4 test sets for Mask-CTC and its extension with intermediate CTC.}
    \vspace{0.5mm}
    \label{tb:maskctc_interctc}
    \renewcommand{\arraystretch}{0.95}
    \scalebox{0.95}{
    \begin{tabular}{lcccc}
        \toprule
        \multirow{2}{*}[-3pt]{\textbf{Model}} & \multicolumn{2}{c}{\textbf{LS-100}} & \multicolumn{2}{c}{\textbf{CHiME4}} \\
        \cmidrule(l{0.2em}r{0.4em}){2-3}
        \cmidrule(l{0.2em}r{0.4em}){4-5}
        & clean & other & real & simu \\
        \midrule
        Mask-CTC & 7.5 & 20.6 & 24.9 & 25.8 \\
        \hspace{1mm} + Intermediate CTC & \textbf{7.2} & \textbf{20.4} & \textbf{24.9} & \textbf{25.0} \\
        \bottomrule
    \end{tabular}}
    \vspace{-1mm}
\end{table}
As intermediate CTC is a regularization method for CTC modeling and does not require any architectural changes, it is possible to augment other CTC-based NAR modeling with intermediate CTC.
We combine Mask-CTC and intermediate CTC by adding an intermediate CTC loss to the encoder of Mask-CTC, as proposed in~\cite{lee2021intermediate}.
Table~\ref{tb:maskctc_interctc} shows the comparison of Mask-CTC and its intermediate CTC extension on LS-100 and CHiME-4.
In all cases, intermediate CTC improved Mask-CTC, similar to experimental results reported by~\cite{lee2021intermediate}, while requiring no extra computation during inference.

\subsection{Application to end-to-end speech translation (E2E-ST)}
\begin{table}[t]
    \centering
    \caption{BLEU ($\uparrow$) scores of speech translation models on Fisher-CallHome Spanish. \#iter is the number of iterations.}\label{tb:result_speech_translation}
    \renewcommand{\arraystretch}{0.95}
    \scalebox{0.94}{
    \begin{tabular}{lccccc}
    \toprule
    \multirow{2}{*}[-3pt]{\textbf{Model}} & \multicolumn{3}{c}{\textbf{Fisher}} & \multicolumn{2}{c}{\textbf{CallHome}} \\
    \cmidrule(lr){2-4} \cmidrule(lr){5-6}
        & dev & dev2 & test & \ devtest \ & evltest \\ \midrule
        CTC & 51.0 & 51.6 & 50.8 & 18.0 & 18.7 \\
        Mask-CTC & 51.1 & 51.7 & 50.6 & 17.9 & 18.3 \\
        Intermediate CTC & 51.3 & 51.4 & \bf{51.0} & 19.0 & 19.0 \\
        Self-conditioned CTC & 50.7 & 51.2 & 50.5 & 19.1 & 19.2 \\
        Orthros ($\text{\#iter}=4$) & 50.1 & 50.6 & 48.7 & 19.5 & 19.8 \\
        Orthros ($\text{\#iter}=10$) & \bf{51.6} & \bf{52.4} & 50.5 & \bf{20.5} & \bf{20.7} \\ 
        \bottomrule
    \end{tabular}}
    \vspace{-1mm}
\end{table}

We applied Mask-CTC, intermediate CTC, self-conditioned CTC, and Orthros~\cite{inaguma2021orthros} to the NAR E2E-ST task on Fisher-CallHome Spanish corpus~\cite{post2013improved}.
Orthros is based on CMLM but has an additional AR decoder on the same encoder to select the most probable translation among multiple length candidates.
All models used the Conformer encoder and 16k wordpieces as output units. 
The encoder parameters were initialized with a pre-trained AR ASR encoder.
We followed the setting in~\cite{inaguma2021orthros} and applied sequence-level knowledge distillation~\cite{kim-rush-2016-sequence}.
With the help of the NAR modeling methods, 
we observed some gains over the CTC-based results (Table~\ref{tb:result_speech_translation}). 
While self-conditioned CTC was the most effective method for ASR, 
it had smaller impact on the E2E-ST task, and 
Orthros resulted in the best performance. 
Since input-output alignments are not monotonic in this task, 
token-level iterative refinement is required to make the sequence generation conditioned on a translated sequence, 
rather than only depending on acoustic information from input speech. 

\section{Conclusions}
This paper presented a comprehensive study of NAR modeling methods for end-to-end ASR. 
Various NAR models were compared based on different ASR tasks. 
The results provided several interesting findings, 
including the accuracy-speed trade-off and robustness against long-form speech utterances.
We also showed that different NAR techniques could be combined to improve the performance and applied to end-to-end speech translation. 
We believe that the reproducible implementations and recipes used in this paper will
accelerate further NAR research in speech processing. 

\section{Acknowledgement}
This work was partly supported by ASAPP.
This work used the Extreme Science and Engineering Discovery Environment (XSEDE) \cite{towns2014xsede}, which is supported by National Science Foundation grant number ACI-1548562. Specifically, it used the Bridges system \cite{nystrom2015bridges}, which is supported by NSF award number ACI-1445606, at the Pittsburgh Supercomputing Center (PSC). 
The authors would like to thank Linhao Dong and Florian Boyer for helpful discussions.

\newpage
\begin{spacing}{0.84}
\bibliographystyle{IEEEbib}
\bibliography{refs}

\begin{thebibliography}{10}

\bibitem{hinton2012deep}
Geoffrey Hinton, Li~Deng, Dong Yu, George~E Dahl, Abdelrahman Mohamed, Navdeep
  Jaitly, Andrew Senior, Vincent Vanhoucke, Patrick Nguyen, Tara~N Sainath,
  et~al.,
\newblock ``Deep neural networks for acoustic modeling in speech recognition:
  The shared views of four research groups,''
\newblock {\em IEEE Signal Process. Mag.}, vol. 29, no. 6, 2012.

\bibitem{graves2013speech}
Alex Graves, Abdelrahman Mohamed, and Geoffrey Hinton,
\newblock ``Speech recognition with deep recurrent neural networks,''
\newblock in {\em Proc. ICASSP}, 2013, pp. 6645--6649.

\bibitem{graves2014towards}
Alex Graves and Navdeep Jaitly,
\newblock ``Towards end-to-end speech recognition with recurrent neural
  networks,''
\newblock in {\em Proc. ICML}, 2014, pp. 1764--1772.

\bibitem{chorowski2015attention}
Jan~K Chorowski, Dzmitry Bahdanau, Dmitriy Serdyuk, Kyunghyun Cho, and Yoshua
  Bengio,
\newblock ``Attention-based models for speech recognition,''
\newblock in {\em Proc. NeurIPS}, 2015, pp. 577--585.

\bibitem{chan2016listen}
William Chan, Navdeep Jaitly, Quoc Le, and Oriol Vinyals,
\newblock ``Listen, attend and spell: {A} neural network for large vocabulary
  conversational speech recognition,''
\newblock in {\em Proc. ICASSP}, 2016, pp. 4960--4964.

\bibitem{graves2012sequence}
Alex Graves,
\newblock ``Sequence transduction with recurrent neural networks,''
\newblock {\em arXiv preprint arXiv:1211.3711}, 2012.

\bibitem{sutskever2014sequence}
Ilya Sutskever, Oriol Vinyals, and Quoc~V Le,
\newblock ``Sequence to sequence learning with neural networks,''
\newblock in {\em Proc. NeurIPS}, 2014, pp. 3104--3112.

\bibitem{bahdanau2014neural}
Dzmitry Bahdanau, Kyunghyun Cho, and Yoshua Bengio,
\newblock ``Neural machine translation by jointly learning to align and
  translate,''
\newblock in {\em Proc. ICLR}, 2014.

\bibitem{dong2018speech}
Linhao Dong, Shuang Xu, and Bo~Xu,
\newblock ``Speech-{Transformer}: a no-recurrence sequence-to-sequence model
  for speech recognition,''
\newblock in {\em Proc. ICASSP}, 2018, pp. 5884--5888.

\bibitem{kriman2020quartznet}
Samuel Kriman, Stanislav Beliaev, Boris Ginsburg, Jocelyn Huang, Oleksii
  Kuchaiev, Vitaly Lavrukhin, Ryan Leary, Jason Li, and Yang Zhang,
\newblock ``Quartznet: {D}eep automatic speech recognition with 1d time-channel
  separable convolutions,''
\newblock in {\em Proc. ICASSP}, 2020, pp. 6124--6128.

\bibitem{gulati2020conformer}
Anmol Gulati, James Qin, Chung-Cheng Chiu, Niki Parmar, Yu~Zhang, Jiahui Yu,
  Wei Han, Shibo Wang, Zhengdong Zhang, Yonghui Wu, and Ruoming Pang,
\newblock ``Conformer: {C}onvolution-augmented {Transformer} for speech
  recognition,''
\newblock in {\em Proc. Interspeech}, 2020, pp. 5036--5040.

\bibitem{chiu2018state}
Chung-Cheng Chiu, Tara~N Sainath, Yonghui Wu, Rohit Prabhavalkar, Patrick
  Nguyen, Zhifeng Chen, Anjuli Kannan, Ron~J Weiss, Kanishka Rao, Ekaterina
  Gonina, et~al.,
\newblock ``State-of-the-art speech recognition with sequence-to-sequence
  models,''
\newblock in {\em Proc. ICASSP}, 2018, pp. 4774--4778.

\bibitem{luscher2019rwth}
Christoph Lüscher, Eugen Beck, Kazuki Irie, Markus Kitza, Wilfried Michel,
  Albert Zeyer, Ralf Schlüter, and Hermann Ney,
\newblock ``{RWTH ASR} systems for {LibriSpeech}: {H}ybrid vs attention,''
\newblock in {\em Proc. Interspeech}, 2019, pp. 231--235.

\bibitem{karita2019comparative}
Shigeki Karita, Nanxin Chen, Tomoki Hayashi, Takaaki Hori, Hirofumi Inaguma,
  Ziyan Jiang, Masao Someki, Nelson Enrique~Yalta Soplin, Ryuichi Yamamoto,
  Xiaofei Wang, et~al.,
\newblock ``A comparative study on {Transformer} vs {RNN} in speech
  applications,''
\newblock in {\em Proc. ASRU}, 2019, pp. 449--456.

\bibitem{graves2006connectionist}
Alex Graves, Santiago Fern{\'a}ndez, Faustino Gomez, and J{\"u}rgen
  Schmidhuber,
\newblock ``Connectionist temporal classification: {L}abelling unsegmented
  sequence data with recurrent neural networks,''
\newblock in {\em Proc. ICML}, 2006, pp. 369--376.

\bibitem{gu2017non}
Jiatao Gu, James Bradbury, Caiming Xiong, Victor~OK Li, and Richard Socher,
\newblock ``Non-autoregressive neural machine translation,''
\newblock in {\em Proc. ICLR}, 2018.

\bibitem{chen2020non}
Nanxin Chen, Shinji Watanabe, Jesus~Antonio Villalba, Piotr Zelasko, and Najim
  Dehak,
\newblock ``Non-autoregressive transformer for speech recognition,''
\newblock {\em IEEE Signal Process. Lett.}, 2020.

\bibitem{chan2020imputer}
William Chan, Chitwan Saharia, Geoffrey Hinton, Mohammad Norouzi, and Navdeep
  Jaitly,
\newblock ``Imputer: {S}equence modelling via imputation and dynamic
  programming,''
\newblock in {\em Proc. ICML}, 2020, pp. 1403--1413.

\bibitem{higuchi2020mask}
Yosuke Higuchi, Shinji Watanabe, Nanxin Chen, Tetsuji Ogawa, and Tetsunori
  Kobayashi,
\newblock ``Mask {CTC}: Non-autoregressive end-to-end {ASR} with {CTC} and mask
  predict,''
\newblock in {\em Proc. Interspeech}, 2020, pp. 3655--3659.

\bibitem{bai2020listen}
Ye~Bai, Jiangyan Yi, Jianhua Tao, Zhengkun Tian, Zhengqi Wen, and Shuai Zhang,
\newblock ``Listen attentively, and spell once: {W}hole sentence generation via
  a non-autoregressive architecture for low-latency speech recognition,''
\newblock in {\em Proc. Interspeech}, 2020, pp. 3381--3385.

\bibitem{fujita2020insertion}
Yuya Fujita, Shinji Watanabe, Motoi Omachi, and Xuankai Chang,
\newblock ``Insertion-based modeling for end-to-end automatic speech
  recognition,''
\newblock in {\em Proc. Interspeech}, 2020, pp. 3660--3664.

\bibitem{tian2020spike}
Zhengkun Tian, Jiangyan Yi, Jianhua Tao, Ye~Bai, Shuai Zhang, and Zhengqi Wen,
\newblock ``Spike-triggered non-autoregressive {Transformer} for end-to-end
  speech recognition,''
\newblock in {\em Proc. Interspeech}, 2020, pp. 5026--5030.

\bibitem{chi2020align}
Ethan~A Chi, Julian Salazar, and Katrin Kirchhoff,
\newblock ``{Align-Refine}: {N}on-autoregressive speech recognition via
  iterative realignment,''
\newblock in {\em Proc. NAACL-HLT}, 2021, pp. 1920--1927.

\bibitem{libovicky2018end}
Jind{\v{r}}ich Libovick{\'y} and Jind{\v{r}}ich Helcl,
\newblock ``End-to-end non-autoregressive neural machine translation with
  connectionist temporal classification,''
\newblock in {\em Proc. EMNLP}, 2018, pp. 3016--3021.

\bibitem{lee2018deterministic}
Jason Lee, Elman Mansimov, and Kyunghyun Cho,
\newblock ``Deterministic non-autoregressive neural sequence modeling by
  iterative refinement,''
\newblock in {\em Proc. EMNLP}, 2018, pp. 1173--1182.

\bibitem{stern2019insertion}
Mitchell Stern, William Chan, Jamie Kiros, and Jakob Uszkoreit,
\newblock ``Insertion {Transformer}: {F}lexible sequence generation via
  insertion operations,''
\newblock in {\em Proc. ICML}, 2019, pp. 5976--5985.

\bibitem{gu2019levenshtein}
Jiatao Gu, Changhan Wang, and Junbo Zhao,
\newblock ``Levenshtein {Transformer},''
\newblock in {\em Proc. NeurIPS}, 2019, pp. 11181--11191.

\bibitem{ghazvininejad2019mask}
Marjan Ghazvininejad, Omer Levy, Yinhan Liu, and Luke Zettlemoyer,
\newblock ``Mask-predict: {P}arallel decoding of conditional masked language
  models,''
\newblock in {\em Proc. EMNLP-IJCNLP}, 2019, pp. 6114--6123.

\bibitem{saharia2020non}
Chitwan Saharia, William Chan, Saurabh Saxena, and Mohammad Norouzi,
\newblock ``Non-autoregressive machine translation with latent alignments,''
\newblock in {\em Proc. EMNLP}, 2020, pp. 1098--1108.

\bibitem{ma2019flowseq}
Xuezhe Ma, Chunting Zhou, Xian Li, Graham Neubig, and Eduard Hovy,
\newblock ``{FlowSeq}: {N}on-autoregressive conditional sequence generation
  with generative flow,''
\newblock in {\em Proc. EMNLP-IJCNLP}, 2019, pp. 4273--4283.

\bibitem{higuchi2021improved}
Yosuke Higuchi, Hirofumi Inaguma, Shinji Watanabe, Tetsuji Ogawa, and Tetsunori
  Kobayashi,
\newblock ``Improved {Mask-CTC} for non-autoregressive end-to-end {ASR},''
\newblock in {\em Proc. ICASSP}, 2021, pp. 8363--8367.

\bibitem{devlin2019bert}
Jacob Devlin, Ming-Wei Chang, Kenton Lee, and Kristina Toutanova,
\newblock ``{BERT}: {P}re-training of deep bidirectional transformers for
  language understanding,''
\newblock in {\em Proc. NAACL-HLT}, 2019, pp. 4171--4186.

\bibitem{chen2021align}
Nanxin Chen, Piotr {\.Z}elasko, Laureano Moro-Vel{\'a}zquez, Jes{\'u}s
  Villalba, and Najim Dehak,
\newblock ``{Align-Denoise}: {S}ingle-pass non-autoregressive speech
  recognition,''
\newblock in {\em Proc. Interspeech}, 2021.

\bibitem{chan2019kermit}
William Chan, Nikita Kitaev, Kelvin Guu, Mitchell Stern, and Jakob Uszkoreit,
\newblock ``{KERMIT}: {G}enerative insertion-based modeling for sequences,''
\newblock {\em arXiv preprint arXiv:1906.01604}, 2019.

\bibitem{lee2021intermediate}
Jaesong Lee and Shinji Watanabe,
\newblock ``Intermediate loss regularization for {CTC}-based speech
  recognition,''
\newblock in {\em Proc. ICASSP}, 2021, pp. 6224--6228.

\bibitem{nozaki2021relaxing}
Jumon Nozaki and Tatsuya Komatsu,
\newblock ``Relaxing the conditional independence assumption of {CTC}-based
  {ASR} by conditioning on intermediate predictions,''
\newblock in {\em Proc. Interspeech}, 2021.

\bibitem{dong2020cif}
Linhao Dong and Bo~Xu,
\newblock ``{CIF}: {C}ontinuous integrate-and-fire for end-to-end speech
  recognition,''
\newblock in {\em Proc. ICASSP}, 2020, pp. 6079--6083.

\bibitem{watanabe2018espnet}
Shinji Watanabe, Takaaki Hori, Shigeki Karita, Tomoki Hayashi, Jiro Nishitoba,
  Yuya Unno, Nelson {Enrique Yalta Soplin}, Jahn Heymann, Matthew Wiesner,
  Nanxin Chen, Adithya Renduchintala, and Tsubasa Ochiai,
\newblock ``{ESPnet}: {E}nd-to-end speech processing toolkit,''
\newblock in {\em Proc. Interspeech}, 2018, pp. 2207--2211.

\bibitem{yu2021boundary}
Fan Yu, Haoneng Luo, Pengcheng Guo, Yuhao Liang, Zhuoyuan Yao, Lei Xie,
  Yingying Gao, Leijing Hou, and Shilei Zhang,
\newblock ``Boundary and context aware training for {CIF}-based
  non-autoregressive end-to-end {ASR},''
\newblock {\em arXiv preprint arXiv:2104.04702}, 2021.

\bibitem{battenberg2017exploring}
Eric Battenberg, Jitong Chen, Rewon Child, Adam Coates, Yashesh Gaur~Yi Li,
  Hairong Liu, Sanjeev Satheesh, Anuroop Sriram, and Zhenyao Zhu,
\newblock ``Exploring neural transducers for end-to-end speech recognition,''
\newblock in {\em Proc. ASRU}, 2017, pp. 206--213.

\bibitem{tjandra2020deja}
Andros Tjandra, Chunxi Liu, Frank Zhang, Xiaohui Zhang, Yongqiang Wang, Gabriel
  Synnaeve, Satoshi Nakamura, and Geoffrey Zweig,
\newblock ``Deja-vu: {D}ouble feature presentation and iterated loss in deep
  transformer networks,''
\newblock in {\em Proc. ICASSP}, 2020, pp. 6899--6903.

\bibitem{ng2021pushing}
Edwin~G Ng, Chung-Cheng Chiu, Yu~Zhang, and William Chan,
\newblock ``Pushing the limits of non-autoregressive speech recognition,''
\newblock in {\em Proc. Interspeech}, 2021.

\bibitem{majumdar2021citrinet}
Somshubra Majumdar, Jagadeesh Balam, Oleksii Hrinchuk, Vitaly Lavrukhin, Vahid
  Noroozi, and Boris Ginsburg,
\newblock ``Citrinet: {C}losing the gap between non-autoregressive and
  autoregressive end-to-end models for automatic speech recognition,''
\newblock {\em arXiv preprint arXiv:2104.01721}, 2021.

\bibitem{baevski2020wav2vec}
Alexei Baevski, Yuhao Zhou, Abdelrahman Mohamed, and Michael Auli,
\newblock ``wav2vec 2.0: {A} framework for self-supervised learning of speech
  representations,''
\newblock in {\em Proc. NeurIPS}, 2020.

\bibitem{vincent2008extracting}
Pascal Vincent, Hugo Larochelle, Yoshua Bengio, and Pierre-Antoine Manzagol,
\newblock ``Extracting and composing robust features with denoising
  autoencoders,''
\newblock in {\em Proc. ICML}, 2008, pp. 1096--1103.

\bibitem{watanabe2020espnet}
Shinji Watanabe, Florian Boyer, Xuankai Chang, Pengcheng Guo, Tomoki Hayashi,
  Yosuke Higuchi, Takaaki Hori, Wen-Chin Huang, Hirofumi Inaguma, Naoyuki Kamo,
  et~al.,
\newblock ``The 2020 {ESPNet} update: {New} features, broadened applications,
  performance improvements, and future plans,''
\newblock in {\em 2021 IEEE Data Science and Learning Workshop (DSLW)}, 2021,
  pp. 1--6.

\bibitem{watanabe2017hybrid}
Shinji Watanabe, Takaaki Hori, Suyoun Kim, John~R Hershey, and Tomoki Hayashi,
\newblock ``Hybrid {CTC}/attention architecture for end-to-end speech
  recognition,''
\newblock {\em IEEE Journal of Selected Topics in Signal Processing}, vol. 11,
  no. 8, pp. 1240--1253, 2017.

\bibitem{karita2019improving}
Shigeki Karita, Nelson Enrique~Yalta Soplin, Shinji Watanabe, Marc Delcroix,
  Atsunori Ogawa, and Tomohiro Nakatani,
\newblock ``Improving {Transformer}-based end-to-end speech recognition with
  connectionist temporal classification and language model integration,''
\newblock in {\em Proc. Interspeech}, 2019, pp. 1408--1412.

\bibitem{guo2021recent}
Pengcheng Guo, Florian Boyer, Xuankai Chang, Tomoki Hayashi, Yosuke Higuchi,
  Hirofumi Inaguma, Naoyuki Kamo, Chenda Li, Daniel Garcia-Romero, Jiatong Shi,
  et~al.,
\newblock ``Recent developments on {ESPnet} toolkit boosted by {Conformer},''
\newblock in {\em Proc. ICASSP}, 2021, pp. 5874--5878.

\bibitem{panayotov2015librispeech}
Vassil Panayotov, Guoguo Chen, Daniel Povey, and Sanjeev Khudanpur,
\newblock ``Librispeech: An {ASR} corpus based on public domain audio books,''
\newblock in {\em Proc. ICASSP}, 2015, pp. 5206--5210.

\bibitem{rousseau2014enhancing}
Anthony Rousseau et~al.,
\newblock ``Enhancing the {TED}-{LIUM} corpus with selected data for language
  modeling and more {TED} talks,''
\newblock in {\em Porc. LREC}, 2014, pp. 3935--3939.

\bibitem{maekawa2000spontaneous}
Kikuo Maekawa, Hanae Koiso, Sadaoki Furui, and Hitoshi Isahara,
\newblock ``Spontaneous speech corpus of {Japanese}.,''
\newblock in {\em Proc. LREC}, 2000, pp. 1--5.

\bibitem{kudo2018subword}
Taku Kudo,
\newblock ``Subword regularization: Improving neural network translation models
  with multiple subword candidates,''
\newblock in {\em Proc. ACL}, 2018.

\bibitem{vincent2017analysis}
Emmanuel Vincent, Shinji Watanabe, Aditya~Arie Nugraha, Jon Barker, and Ricard
  Marxer,
\newblock ``An analysis of environment, microphone and data simulation
  mismatches in robust speech recognition,''
\newblock {\em Computer Speech \& Language}, vol. 46, pp. 535--557, 2017.

\bibitem{paul1992design}
Douglas~B Paul and Janet~M Baker,
\newblock ``The design for the wall street journal-based {CSR} corpus,''
\newblock in {\em Proc. Workshop on Speech and Natural Language}, 1992, pp.
  357--362.

\bibitem{povey2011kaldi}
Daniel Povey, Arnab Ghoshal, Gilles Boulianne, Lukas Burget, Ondrej Glembek,
  Nagendra Goel, Mirko Hannemann, Petr Motlicek, Yanmin Qian, Petr Schwarz,
  et~al.,
\newblock ``The {Kaldi} speech recognition toolkit,''
\newblock in {\em Proc. ASRU}, 2011.

\bibitem{ko2015audio}
Tom Ko, Vijayaditya Peddinti, Daniel Povey, and Sanjeev Khudanpur,
\newblock ``Audio augmentation for speech recognition,''
\newblock in {\em Proc. Interspeech}, 2015, pp. 3586--3589.

\bibitem{park2019specaugment}
Daniel~S Park, William Chan, Yu~Zhang, Chung-Cheng Chiu, Barret Zoph, Ekin~D
  Cubuk, and Quoc~V Le,
\newblock ``{SpecAugment}: {A} simple data augmentation method for automatic
  speech recognition,''
\newblock in {\em Proc. Interspeech}, 2019, pp. 2613--2617.

\bibitem{kingma2015adam}
Diederik~P Kingma and Jimmy Ba,
\newblock ``Adam: A method for stochastic optimization,''
\newblock in {\em Proc. ICLR}, 2015.

\bibitem{vaswani2017attention}
Ashish Vaswani, Noam Shazeer, Niki Parmar, Jakob Uszkoreit, Llion Jones,
  Aidan~N Gomez, {\L}ukasz Kaiser, and Illia Polosukhin,
\newblock ``Attention is all you need,''
\newblock in {\em Proc. NeurIPS}, 2017, pp. 5998--6008.

\bibitem{chorowski2017towards}
Jan Chorowski and Navdeep Jaitly,
\newblock ``Towards better decoding and language model integration in sequence
  to sequence models,''
\newblock in {\em Proc. Interspeech}, 2017, pp. 523--527.

\bibitem{inaguma2021orthros}
Hirofumi Inaguma, Yosuke Higuchi, Kevin Duh, Tatsuya Kawahara, and Shinji
  Watanabe,
\newblock ``Orthros: {N}on-autoregressive end-to-end speech translation with
  dual-decoder,''
\newblock in {\em Proc. ICASSP}, 2021, pp. 7503--7507.

\bibitem{post2013improved}
Matt Post, Gaurav Kumar, Adam Lopez, Damianos Karakos, Chris Callison-Burch,
  and Sanjeev Khudanpur,
\newblock ``Improved speech-to-text translation with the {Fisher} and {Callhome
  Spanish--English} speech translation corpus,''
\newblock in {\em Proc. IWSLT}, 2013.

\bibitem{kim-rush-2016-sequence}
Yoon Kim and Alexander~M. Rush,
\newblock ``Sequence-level knowledge distillation,''
\newblock in {\em Proc. EMNLP}, 2016, pp. 1317--1327.

\bibitem{towns2014xsede}
John Towns, Timothy Cockerill, Maytal Dahan, Ian Foster, Kelly Gaither, Andrew
  Grimshaw, Victor Hazlewood, Scott Lathrop, Dave Lifka, Gregory~D Peterson,
  et~al.,
\newblock ``{XSEDE}: Accelerating scientific discovery,''
\newblock {\em Computing in Science \& Engineering}, vol. 16, no. 5, pp.
  62--74, 2014.

\bibitem{nystrom2015bridges}
Nicholas~A Nystrom, Michael~J Levine, Ralph~Z Roskies, and J~Ray Scott,
\newblock ``Bridges: a uniquely flexible {HPC} resource for new communities and
  data analytics,''
\newblock in {\em Proc. the 2015 XSEDE Conference: Scientific Advancements
  Enabled by Enhanced Cyberinfrastructure}, 2015, pp. 1--8.

\end{thebibliography}
\end{spacing}

\end{document}